# «REGULATING AI: DO WE NEED NEW TOOLS?»

Otello Ardovino[(*)], Autorità per le Garanzie nelle Comunicazioni
Jacopo Arpetti, University of Rome Tor Vergata
Marco Delmastro[(*)], Autorità per le Garanzie nelle Comunicazioni

**Abstract**

The Artificial Intelligence paradigm (hereinafter referred to as "AI") builds on the analysis of data able, among other things, to snap pictures of the individuals' behaviors and preferences. Such data represent the most valuable currency in the digital ecosystem, where their value derives from their being a fundamental asset in order to train machines with a view to developing AI applications. In this environment, online providers attract users by offering them services for free and getting in exchange data generated right through the usage of such services. This swap, characterized by an implicit nature, constitutes the focus of the present paper, in the light of the disequilibria, as well as market failures, that it may bring about. We use mobile apps and the related permission system as an ideal environment to explore, via econometric tools, those issues. The results, stemming from a dataset of over one million observations, show that both buyers and sellers are aware that access to digital services implicitly implies an exchange of data, although this does not have a considerable impact neither on the level of downloads (demand), nor on the level of the prices (supply). In other words, the implicit nature of this exchange does not allow market indicators to work efficiently. We conclude that current policies (e.g. transparency rules) may be inherently biased and we put forward suggestions for a new approach.

**Contacts**:

Otello Ardovino, Department of Economics & Statistics, corresponding author (http://orcid.org/0000-0001-9226-6348)
Autorità per le Garanzie nelle Comunicazioni - Centro Direzionale, Isola B5 - 80143, Naples, Italy
E-mail: o.ardovino@agcom.it

Jacopo Arpetti, corresponding author (http://orcid.org/0000-0002-3448-1055)
Department of Enterprise Engineering, University of Rome Tor Vergata
Via del Politecnico, 1 - 00133, Rome, Italy
E-mail: jacopo.arpetti@uniroma2.it

Marco Delmastro, Department of Economics & Statistics, corresponding author (http://orcid.org/0000- 0002-8527-3117)
Autorità per le Garanzie nelle Comunicazioni - Via Isonzo, 21/b - 00198, Rome, Italy
E-mail: m.delmastro@agcom.it

[(*)] The authors gratefully acknowledge Andrea Vitaletti for helpful comments. The usual disclaimer applies. The views expressed herein by Otello Ardovino and Marco Delmastro are the sole responsibility of the author and cannot be interpreted as reflecting those of the Autorità per le Garanzie nelle Comunicazioni.

# 1. Introduction

The whole AI paradigm builds on the analysis of data mostly generated by individuals and then used to train machines (supervised, semi-supervised and unsupervised machine learning). Data stemming from the individuals' behaviors and preferences (be it of a personal nature or not) therefore represent one of the most valuable "currency" in a data-hungry digital ecosystem featured to a large extent by AI applications.

A problem arises relating to the collection of such data, which revolves around the implicit nature of transactions involving, on the one hand, services offered by platforms and, on the other hand, data ceded by individuals. The present contribution focuses on this relation that shapes market outcomes, in terms of economic and social (static and dynamic) efficiency.

The emerging digital economy is characterized by a "data-driven" business model (Delmastro & Nicita, 2019), which is devised to create value via data aggregation and analysis which, in turns, are made possible by the individuals' choice to cede their own data (or to allow their collection) in exchange for services offered by platforms (see the definition below). The whole model therefore relies upon the implicit swap of data for services and this, in addition to a wide array of (positive and negative) externalities and lock-in effects for the consumer, can result into further market failures.

With reference to the mentioned data-related transactions, it is worth noting that some literature addresses indeed privacy not as an absolute right of the individual, but rather as a "sphere" subject to economic dynamics; in this sense, privacy can be conceived as a commodity, implying relevant trade-offs and therefore encompassing cost-benefit evaluations by the individual (for the notion of privacy as a commodity, see Bennett; Davies (Bennett, 1995; Davies, 1997)).

Although any data transfer can theoretically be equaled to an exchange of goods in terms of relevant dynamics and individual's assessments, it is nonetheless not accurate to place such transactions on the same level, as it is extremely difficult in practice for the individual to determine the real economic value of the data they provide.

The above-mentioned approach to privacy as a commodity has triggered a heated debate [see (Cohen, 2000)] with respect to the notion of individuals' bounded rationality. As a matter of fact, when individuals provide their consent to transfer their data, they perform an evaluation in terms of cost-benefit, similar to the one carried out in relation to every purchase decision. However, when making decisions concerning whether providing their own data to an online subject, hence when defining their digital behaviors (i.e. consuming or not a digital service vis à vis a certain privacy policy being applied by online players) the individuals' reasoning is aimed at understanding whether it would be convenient to provide their personal information in exchange for benefits of another nature (economic or not). Such weighing in exercise is carried out in a context where users of online digital services do not avail of all the information necessary to measure the costs they will actually bear (uncertain as well as potential ones) because the environment in which they are immersed makes it difficult to thoroughly carry out evaluations as such.

In this context, furthermore characterized by transactions that are not explicitly evident to digital users (i.e. "implicit transactions"), evaluations on the exchange of data for services of other nature are affected by the individuals' limitations in terms of capacity to define the marginal value of the non-monetary, incremental benefits «in relation to the focal product or its price» (Acquisti & Grossklags, 2005, 2008b). Such constraints, sharpened by incomplete information, information asymmetry and bounded rationality (Akerlof, 1970; Arrow, 1958; Simon, 1955), force individuals to conduct evaluations about the outcomes of data transfers in exchange for non-monetary and often intangible assets (mostly services) in a context marked by uncertainty and a high level of complexity. This affects the individuals' assessment of the relevant consequences, hence their probability of occurrence «since the states of nature may be unknown or unknowable in advance» (Acquisti & Grossklags, 2008a).

The digital environment is featured indeed by exogenous and endogenous components which make an evaluation difficult, starting with uncertainty, which constitutes the first element influencing the individuals' choice. Together with their own preferences in relation to the data transfer itself, individuals are indeed called to cope with the uncertain nature of the trade-offs implied by online transactions.

Moreover, due to the technological progress and to the ever more pervasive data collection practices implemented, asymmetric information further increases. As a matter of fact, the emergence of the data economy paradigm makes data collection almost imperceptible – if not invisible – to the user, who therefore has limited awareness of how much data is being collected about him/her, what kind of data he/she is actually ceding to platforms in order to access certain services, how the acquired data will be used and with what consequences. This recalls the Acquisti "blank check" metaphor (Acquisti, 2010, p. 15,16).

The remainder of the paper is organized as follows. In Section 2, we provide a conceptual framework, which constitutes the backdrop to the analysis. In Section 3, we illustrate the study design of our empirical analysis aimed at studying the (implicit) transactions between consumers and online operators. In Section 4, we present the results from econometric models' elaborations, relying on millions of data on mobile apps. Lastly, Section 5 provides our conclusions together with potential policy implications.

## 2. Conceptual framework

Setting aside for the moment the non-proper transactional nature of the relations at stake, the context in which individuals adopt their own data-related decisions in exchange for online services is shaped by information asymmetry and bounded rationality.

The traditional landmark trait describing rationality in economic terms was long considered to be the consumers' goal of maximizing their expected utility. In the wake of the introduction of the notion of "bounded rationality", Kahneman and Tversky have shown instead logical inconsistencies in individuals' preferences with respect to the axiom of rationality, showing how preferences are influenced by framing, hence by the alternative options' formulation (Kahneman & Tversky, 1986), and then by *reference dependence* and *prospect theory* (Kahneman, 2011; Kahneman & Tversky, 1979; Tversky & Kahneman, 1991).

With respect to *reference dependence*, Kahneman and Tversky have shown how individuals evaluate outcomes taking a point as a refence (*reference point*) in order to classify gains and losses (Tversky & Kahneman, 1991). Starting from this assumption, Adjerid, Acquisti, Brandimarte, and Loewenstein (2013), providing individuals with information about the subsequent use of their data acquired in a digital environment, have shown that their preferences concerning data transfer can change, to the point that identical privacy notices do not always lead to the same level of data disclosure. Plunged in a universe featured by elements to which they cannot assign a value, individuals appear unable to figure out the probability of occurrence of a given event, nor the consequences of their actions.

Starting from the '80s, numerous theories have shown – in different contexts – a deviation from the then consolidated economic concept of rationality [e.g. (Benartzi & Thaler, 1995)]. In our paper, we make reference to the digital dynamic of data ceded by individuals in exchange for services of other nature, in order to pinpoint the by now nuanced concept of rationality in the individuals' economic choices and to accordingly show how individuals make "no rational" choices in the data society context.

In the digital environment, individuals' choices on data transfer are affected indeed by incomplete and asymmetric information (Akerlof, 1970) due to the fact that online digital users have often no idea of the amount of data really gathered by online players, nor of the way how such data will be used, or to whom they could be sold. Individuals are not in a position to foresee the short-term uses, nor the prospective treatments which data ceded by them could be subject to.

Due to the complexity inherent to the current information society, and to the mentioned individuals' bounded rationality, data subjects are forced to make use of simplified models to decrypt the digital reality, while the presence of heuristics would challenge rational decision-making logics also in a hypothetical condition of complete information (in the light of cognitive anomalies).

Furthermore, the emergence of a new awareness in digital users with respect to the intrinsic value of their data and to data ceding-related risks remains weak, due to the perceived underlying non-transactional nature of the data transfer. Due to the implicit nature of digital transactions in an online environment, individuals do not manage to assign an economic value to the data that they are about to cede, let alone to compare them with services obtained in return, or to weigh up the risks connected to their choices, assigning a probability of occurrence to their consequences. The nature of the elements involved in such "swap" therefore differs to the point that individuals cannot attach a proper value to their data, as they do not perceive the transactional nature of the situation they experience. Having no alternatives or valid reference points to break up this "loop", individuals impulsively cede their data, accepting almost any conditions imposed on them by platforms, which already amounts to a market failure.

If uncertainty related to incomplete information and information asymmetry could lead to a low risk perception (again, considering how both bounded rationality and cognitive anomalies could affect choices also in a context characterized by complete information), it is possible to affirm that, given the impulsive choices adopted when they exchange data ignoring some of the variables that come into play, individuals share some of the biases with subjects affected by "gambler's fallacy", jumping to hasty conclusions and making a bet on their data, thinking that transactions would always turn to their advantage.

The present contribution intends to analyze how all the elements featuring the constrained individual's rationality, together with the misperception concerning the transactional nature of end-users' online digital choices as to the use of their data, drive to inefficient market outcomes.

We study this relation in an empirical setting – i.e. mobile apps and the related permission system – which provides us with an ideal testbed. As a matter of fact, the apps permission system is currently the most transparent and standardized form of data exchange in the digital ecosystem. Indeed, permissions, on the one hand, inform consumers on the nature of individual data that will be gathered by online operators (see infra) and, on the other hand, they represent the finalization of a transaction between the two parties (data and money in exchange for online services). Of course, this system does not provide all the information needed by consumers with a view to making their decisions. However, it constitutes an ideal benchmark to test our framework. In fact, all the other systems are less informative, so that results from this "natural experiment" are inherently biased towards a more efficient outcome. In other words, we study the data exchange relation between consumers and online operators by placing ourselves in the worst possible condition to verify the existence of market failures. In the next paragraph, we illustrate our empirical setting in detail.

## 3. Empirical Analysis

### 3.1 Study design

As already said, the aim of our research is to analyze the data transactions between consumers and digital operators. To this end, we focus on the apps permission system. As a matter of fact, the permission system is a formal environment which offers a framework within which individuals cede data in exchange for digital services (i.e. mobile apps).

All the findings about apps permissions which are proposed in the present paper are based on a dataset containing data from about 1,135,700 apps (offered on the Android Google Play Store) which have been collected through crawling techniques (see also AGCOM, 2018)[1].

Permissions can be clustered based on the amount and nature of the costumers' information they collect. In this sense, we can distinguish between permissions that can have an impact in the wider terms of data collection in general (they would gather whatever data stemming from users' behaviors), and permissions which require access and usage of the user's digital data in order to ensure the proper functioning of the application.

In the light of such differentiation, the present paper makes reference to permissions categorizations emerged in part of the literature. In this respect, the Pew Research Center (2014) groups apps permissions into two categories: those which access to device hardware and those that access to various types of user information[2]. In particular, the latter allow to gather a greater amount of individual data, which are often irrelevant with a view to the proper functioning of the app. Another interesting classification is the one proposed by Kummer and Schulte (2016), who – building on a previous classification put forward by Sarma et al. (2012) – identify a limited number of apps permissions as critical in terms of sensitive data. Finally, we use a more technical classification adopted by Google itself.[3]

---

[1] Crawler architecture refers to the copying process of web contents (data) following links to reach numerous pages through which it was possible to gather information on apps present on the Android Google Play Store. The collection of data and information was carried out by the Department of Computer, Control, and Management Engineering, University "La Sapienza", Rome We gratefully acknowledge A. Vitaletti and A. De Carolis (see also Agcom, 2018).

[2] Pew Research Center (2015), *Apps Permissions in the Google Play Store*, www.pewinternet.org

Referring to the 10 most widespread apps permissions, Table 1 summarizes whether these permissions are considered as sensitive – according to the Pew Research and Kummar and Schulte classifications – in terms of digital footprints left by the users in the virtual environment, and whether they should be considered as dangerous or normal according to the Google classification.

**Table 1: Classifications of apps permissions with a view to individual data**

| Permission type and name | The permission… | Pew Center | Kummer & Schulte | *Google* |
|---|---|---|---|---|
| *full network access* <br><br> *"INTERNET"* | Allows applications to create network sockets and use custom network protocols. | Yes | No | Normal |
| *view network connections* <br><br> *"ACCESS_NETWORK_STATE"* | Allows applications to access information about network connections such as which network exist and are connected. | No | No | Normal |
| *Read exsternal storage* <br><br> "READ_EXTERNAL_STORAGE" <br><br> "WRITE_EXTERNAL_STORAGE" | Allows an app to access to read from an external storage. | Yes | No | Dangerous |
| *Write external storage* | Allows an app to access to write on an external storage. Any app that declares the "write external storage" permission is implicitly granted the permission to read also. | Yes | No | Dangerous |
| *read phone status and identity* <br><br> *Permission group: "PHONE", which includes:* <br><br> • *"READ_PHONE_STATE"*; <br> • *"READ_PHONE_NUMBERS"*; <br> • *"CALL_PHONE"*; <br> • *"ANSWER_PHONE_CALLS"*; <br> • *"ADD_VOICEMAIL"*; <br> • *"USE_SIP"*. | group allows access to the device identifiers. "<u>Read_phone_state</u>", e.g., allows the access to the phone state, including the phone number of the device, current cellular network information, the status of any ongoing calls, and a list of any PhoneAccounts. registered on the device | Yes | Yes | Dangerous |
| *prevent device from sleeping* <br><br> *"WAKE_LOCK"* | Allows using *PowerManager WakeLocks* to prevent processor from sleeping or screen from dimming. | No | No | Normal |
| *view WI-FI connections* <br><br> *"ACCESS_WIFI_STATE"* | Allows applications to access information about Wi-Fi networks. | Yes | No | Normal |
| *precise location GPS and network-based* <br><br> *"ACCESS_FINE_LOCATION"* | Allows an app to access precise location access precise location from location | Yes | Yes | Dangerous |

---

[3] For further information see: https://developer.android.com/guide/topics/permissions/overview

| | sources such as GPS, cell towers, and Wi-Fi. | | | |
|---|---|---|---|---|
| *control vibration* "VIBRATE" | Allows access to the vibrator. | No | No | Normal |
| *approximate location network-based* "ACCESS_COARSE_LOCATION" | Allows an app to access approximate location derived from network location sources such as cell towers and Wi-Fi. | Yes | Yes | Dangerous |

The 1,135,700 apps composing the dataset overall contain 266 "unique" permissions.[4] It should also be stressed that a considerable number of permissions, as described above, refers to technical aspects enabling the proper functioning of the apps. As an example, should a developer want to design a mapping app, it would be necessary to provide this with the right permissions, hence ensuring its access to GPS sensors-related data on the device on which the app will be installed.

Figure 1 shows the permissions distribution among apps: only 10 out of 266 permissions are used by more than 20% of the surveyed apps. A considerable amount of permissions is therefore used by few applications; 20 out of these permissions are used only by 20 apps. The "long-tail" theory could be also applied to the permissions' distribution (Anderson, 2006).

**Figure 1: Permission distribution**

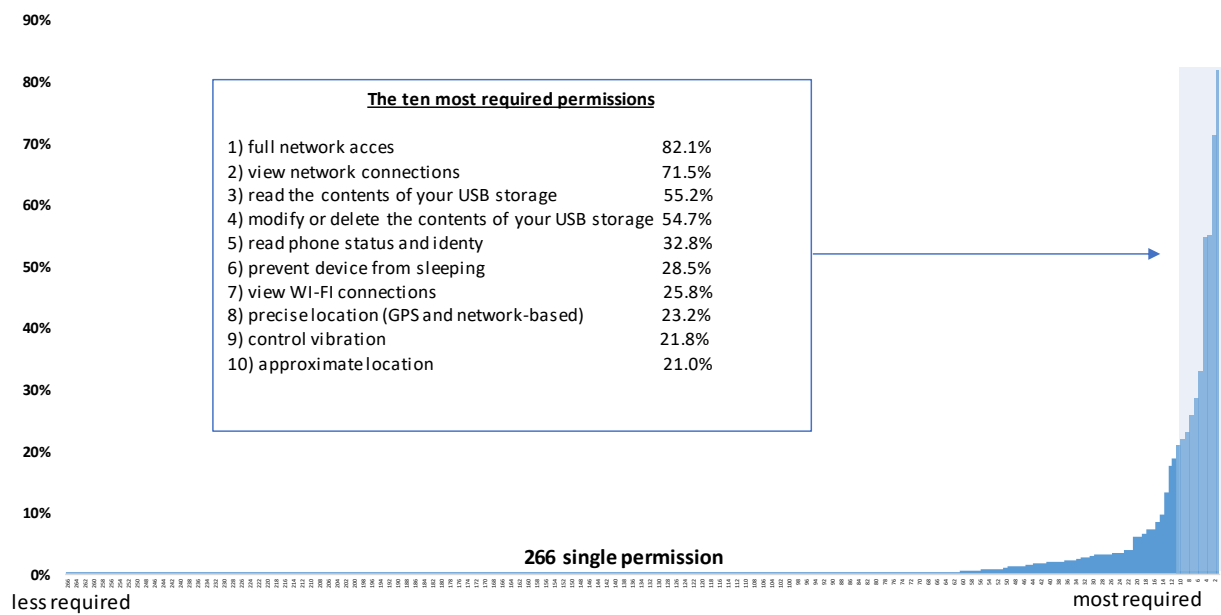

The ten most required permissions
1) full network acces  82.1%
2) view network connections  71.5%
3) read the contents of your USB storage  55.2%
4) modify or delete the contents of your USB storage  54.7%
5) read phone status and identy  32.8%
6) prevent device from sleeping  28.5%
7) view WI-FI connections  25.8%
8) precise location (GPS and network-based)  23.2%
9) control vibration  21.8%
10) approximate location  21.0%

266 single permission
less required — most required

---

[4] Android's development platform, the platform for developing apps for the Google Play Store, allows developers to envisage also new types of permissions or to replace some of them. In such case, new aspects linked to sector evolution might come up (emerging technologies, new types of apps), which might prompt the need for new types of permissions.

## 3.2 Qualitative findings

Descriptive statistics provide useful preliminary information on the phenomenon at stake. As for the price variable (Table 2), the distribution of apps in the market is extremely asymmetric: 86% of them can indeed be downloaded for free, while only 0.5% – precisely 5.171 apps – has a price above 10 euros.

Table 2: apps distribution by price range

|  | Number of apps | % |
|---|---|---|
| free | 977,244 | 86.05 |
| price>0 | 158,456 | 13.95 |
| *0-0.99 €* | *65.676* | *5.8* |
| *1-1.99 €* | *46.882* | *4.1* |
| *2-4.99 €* | *33.415* | *2.9* |
| *5-9.99 €* | *7.312* | *0.6* |
| *≥10€* | *5.171* | *0.5* |
| **Total** | **1,135,700** | **100** |

The fact that an app can be downloaded for free does not prevent the user from deciding, at a following stage, to take advantage of the in-app purchase service, allowing him/her to obtain additional services (so called *in-app purchase*), upon payment of a given amount of money.

Comparing the average number of permissions between free and paid apps (i.e. Table 3), it emerges that free apps require a statistically significant higher number of permissions (on average, 6.4) than the paid ones (on average, 3.8).

Table 3: average number of permissions, a comparison between paid and free apps

|  | # of apps | % | average # of permissions |
|---|---|---|---|
| Free apps | 977,244 | 86% | 6.4* |
| Paid apps | 158,456 | 14% | 3.8* |
| **Total apps** | **1,135,700** | **100%** | **6.0** |
| *t-test, significant at 1% | | | |

Considering the three categories previously outlined (according to the Pew Research, Kummar-Schulte and Google classifications), the results do not change. Both in the case of apps that require at least one sensitive permission, and in all other cases, the average number of requested permissions is markedly higher when the apps can be downloaded for free. As a consequence, the provision of an app for free implies the ceding, through the user's acceptance of the conditions outlined in the permissions, of a greater amount of digital data, particularly those regarding users' sensitive information. In turns, an implicit exchange emerges between users and internet platforms, that affects the primary commercial relationship concerning the purchase and sale of apps, and that implies that the free access to an app is associated to the user's consent to a higher number of permissions, particularly to those related to the disclosure of personal data.

In addition, it emerges that the most downloaded apps are those featuring a higher number of permissions; apps showing over 100,000 downloads require on average 2.5 more permissions compared to apps downloaded between 10,001 and 100,000 times, and 3 more permissions compared to apps downloaded between 1001 and 10,0000 times. Considering apps downloaded over 100 million times, the average number of their permissions rises to 20, slightly less than 4 times the average amount (6.0 permissions).

The number of downloads represents a valid proxy of the users' demand for apps; the distribution of downloads, according to whether the apps are free or not, shows some differences. As it was easy to imagine, free apps show a higher number of downloads. More than 80% of paid apps are indeed downloaded from 1 up to 100 times, whereas for free apps, such figure reaches around 45%[5].

## 4. Econometric models and results

Exploiting econometric tools, we detected some insights about users' (*demand* side) and developers' (*supply* side) reactions to the number and the typologies of permissions required by apps. For both sides, we propose an econometric analysis composed of three models estimated according to the above-mentioned permissions classification; the first one takes into consideration the Pew Research classification; the second one the Kummer and Schulte's one, while the last model adopts the classification implemented by Google.

As concerns the analysis of the apps *demand*, a linear regression (OLS) has been estimated, modelling the demand for apps in terms of downloads as a function of the permissions that they require on a set of control variables for each app such as the price, the category, the average rating, the number of reviews and the app developer.

The econometric model is as follows:

$$Demand_i = \alpha + \beta D_i + \theta X_i + \epsilon_i$$

where the dependent variable (*Demand*) is represented by the logarithm of the total number of downloads for a generic app *i*, $X_i$ includes the control variables, $D_i$ is a dummy variable linked with each permission considered in the proposed classification and, finally, $\epsilon_i$ represent the classic error term.

Parameter $\beta$ is of utmost interest due to its association to the vector of dummy variables that represent each permission; if the estimated parameter appears to be significant and negative, as expected, that implies that the presence of permissions sensitive to individual data leads to a reduction in the demand.

---

[5] The downloads trend shows to what extent the long-tail theory is also valid for the apps market; in fact, setting aside the distinction between paid and non-paid apps, about 50% of the apps is downloaded less than 100 times, and about 98% less than 100,000 times. This shows how just a few apps, i.e. 2% (the long-tail), have been installed by a considerable number of users. A total of 6 apps has been installed more than 1,000,000,000 times; Facebook, Google Gmail, YouTube, Google Maps, Google Search and Google Play services, with an average number of permissions equal to 43.5.

The results (Table 3) confirm the presence of a negative and significant direct effect of the number and typologies of permissions on the number of downloads. This relation can be also detected in Model C, in which we consider a more detailed and specific classification of permission, as suggested by Kummer and Schulte.

**Table 4: demand model (OLS models)**

| Dependent variable: log. of installations | Model A: *Pew Research classification* | Model B: *Google classification* | Model C: *Kummer and Schulte* |
|---|---|---|---|
| User information permissions | -0.05*** | | |
|  | (0.00) | | |
| Dangeorus permissions | | -0.01*** | |
|  | | (0.00) | |
| Full Internet access permissions | | | 0.07*** |
|  | | | (0.00) |
| View network state permissions | | | -0.13*** |
|  | | | (0.00) |
| Phone state permissions (read phone state and ID) | | | 0.07*** |
|  | | | (0.00) |
| Location permissions (Gps) | | | -0.05*** |
|  | | | (0.00) |
| Communication permissions (read sms, intercept outgoing calls, ecc.) | | | -0.10*** |
|  | | | (0.00) |
| Users profile permissions | | | -0.02*** |
|  | | | (0.00) |
| Other permissions | | | -0.06*** |
|  | | | (0.01) |
| Constant | 1.95*** | 1.92*** | 2.00*** |
|  | (0.02) | (0.02) | (0.03) |
| Controls | Yes | Yes | Yes |
| Categories | Yes | Yes | Yes |
| Adjusted $R^2$ | 0.84 | 0.84 | 0.84 |
| # of observations | 1,135,700 | 1,135,700 | 1,135,700 |

Heteroskedasticity-robust standard errors in brackets. ***, **, * significantly different from 0 at 1%, 5% and 10% levels, respectively

With regards to the analysis of the supply side, we focus on apps price; as seen above, the descriptive statistics show a clear difference in the average number of permissions between paid apps and free apps. With the econometric analysis, we provide rigorous evidence in support of the assumption that business models adopted by developers are systematically affected by choices concerning the number and type of permission that users would have to accept when downloading the app.

To this purpose, we estimate a probabilistic model in order to take into account the business model adopted by the developers via a dichotomous variable which is equal to 1 if an app is paid and to 0 if not.

The model is as follows:

$$\Pr(Price_i = 1) = \Lambda[\alpha + \beta D_i + \theta X_i + \epsilon_i]$$

Also in this case, the parameter of greatest interest is represented by $\beta$ which, combined with the dummy variable ($D$), identifies whether a permission can be considered as relevant in terms of transfer of individuals' data; $\epsilon_i$ is the classical error term.

The presence of permissions sensitive to individual data requested by an app (among those which collect individuals' data) reduces the likelihood that the application presents a price higher than 0. Table 5 shows the results of the estimation of the three models, in analogy with the analysis conducted on the demand. The results also show that, on the supply side, a significant and negative relationship emerges between the number and the type of permissions and the probability that the app is not offered for free on the app store.

**Table 5: supply model (probit models)**

| Dependent variable: APPs price | Model A: *Pew Research classification* | Model B: *Google classification* | Model C: *Kummer and Schulte* |
|---|---|---|---|
| User information permissions | -0.26*** | | |
|  | (0.00) | | |
| Dangeorus permissions | | -0.67*** | |
|  | | (0.00) | |
| Full Internet access permissions | | | -0.41*** |
|  | | | (0.01) |
| View network state permissions | | | -0.55*** |
|  | | | (0.00) |
| Phone state permissions (read phone state and ID) | | | 0.09*** |
|  | | | (0.00) |
| Location permissions (Gps) | | | -0.30*** |
|  | | | (0.01) |
| Communication permissions (read sms, intercept outgoing calls, ecc.) | | | 0.01 |
|  | | | (0.01) |
| Users profile permissions | | | -0.00 |
|  | | | (0.01) |
| Other permissions | | | -0.03 |
|  | | | (0.02) |
| Constant | -0.33*** | 0.08** | 0.29*** |
|  | (0.03) | (0.03) | (0.04) |
| Controls | Yes | Yes | Yes |
| Categories | Yes | Yes | Yes |
| Adjusted R$^2$ | 0.14 | 0.16 | 0.21 |
| # of observations | 1,135,700 | 1,135,700 | 1,135,700 |

Heteroskedasticity-robust standard errors in brackets. ***, **, * significantly different from 0 at 1%, 5% and 10% levels, respectively

A more interesting result emerges from the interpretation of these outcomes as a whole. In fact, in addition to the significance of the estimates, a weak magnitude effect (or elasticity) of permissions appears, both on the demand and supply side.

These results suggest that, even if consumers were plunged in a hypothetical "nearly perfect market" configuration in which buyers and sellers avail of complete information about a particular product, thus being very easy to compare prices as well as the specific characteristics of different apps (with particular attention to the number and the typology of the permissions required), things would not work as they should. Indeed, both buyers and sellers are aware that access to digital services implicitly implies an exchange of data (significance of the coefficient $\beta$), although this does not have considerable impact (as it happens instead in the classical goods markets) neither on the level of downloads (demand), nor on the level of the prices (supply). In other words, the implicit nature of this exchange does not allow market indicators to work efficiently.

On the one hand, individuals' behavior, as mentioned in previous sections, is affected by bounded rationality. In such context, where a consumer purchase requires relatively low efforts in terms of money, time, physical and mental commitment, and the price - in particular its low level - does not appear to be the decisive factor in consumer choices, impulsive behaviors often materialize (Bayley & Nancarrow, 1998; Rook & Fisher, 1995; Rook & Hoch, 1985; Stern, 1962). The easier to buy a good, especially in terms of price affordability, the higher the chance of making an impulsive purchase, as choices are hedonically complex and more emotional than rational.

On the other hand, in a digital environment, a number of conditions may ease "impulsive" choices by users. Users' online choices are much more versatile than what is suggested by the "rationality" hypothesis, often because online shopping is associated with a hedonic experience, in which purchasing choices have a reduced time horizon (Moe, 2003). These contexts make the user less focused on the purchase decision process, so that this latter appears to be more stimulus-driven than goal-driven.

Such aspect may also explain why, from the app store point of view, the algorithm used to suggest apps to users is based on a set of information (the number of downloads reached, as well as the app rating) that is able to influence impulsive purchases. Therefore, if faced with the right type of stimulus, the probability that a consumer makes an impulse purchase becomes higher. Moreover, while under the rational paradigm the purchase is planned to the smallest details, and often this requires time, in the digital environment, the relevant conditions allow consumers to make a purchase immediately. Many customers are conscious of what they are doing (an impulsive purchase), but this represents the way in which they reconcile utilitarian and hedonic factors affecting the online surfing experience (Akram et al., 2017).

In this paper, however, we put forward another and more structural mechanism favoring impulsive behavior and inefficient market outcomes, i.e. the implicit nature of these transactions. In our viewpoint, it is indeed the very nature of the current functioning of the digital ecosystem, in which no formal data transaction emerges (so that economic indicators, e.g. prices, that normally regulate markets, are ruled out), that prevents the emergence of socially efficient outcomes.

## 5. Policy considerations

In the data-driven context, data are essential to enable AI: those stemming from online services represent one of the most relevant catalysts capable to record and analyze the characteristics of the environment surrounding individuals, as well as the way in which these latter move within it and the interaction between them in a certain dimension, ultimately converting such data into essential assets through which AI systems can be trained.

Indeed, as Tirole affirms, in the digital environment, data have increasingly taken on a value comparable to that of money: «People often argue that platforms should pay for the data we give them. In practice, many sites do pay. This payment does not take the form of a financial transfer, but rather of services provided free of charge. We provide our personal data in exchange either for useful services (search engines, social networks, instant messaging, online video, maps, email) or in the course of a commercial transaction (as in the case of Uber and Airbnb). Online businesses can often argue that they have spent money to acquire our data. […] As data is at the heart of value creation, defining rules governing its use is an urgent task.» (Tirole, 2017, p. 399).

In particular, in this paper we have shown, through an empirical investigation into big data – considered as an essential asset with a view to AI – that market failures are intrinsic to the very nature of digital transactions, and that traditional policies do not therefore apply to the digital context.

As a matter of fact, EU law currently provides consumer protection tools when it comes to B2C relations characterized by the provision of goods or services upon payment of a monetary amount to the platform. Key reference provisions – currently under review – addressing the terms and modalities of the monetary transactions between consumers and undertakings, including platforms, are indeed currently provided by Directive 93/13 on unfair terms in consumer contracts (European Council, 1993), Directive 98/6/EC on consumer protection in the indication of the prices of products offered to consumers, Directive 2005/29 on unfair business-to-consumer commercial practices in the internal market (European Parliament and of the Council, 2005)[6] and Directive 2011/83 on consumer rights (European Parliament and European Council, 2011).

Nevertheless, when consumers obtain, in a digital environment, services in exchange for their data – therefore without a corresponding payment of monetary value – the mentioned provisions do not provide complete protection to individuals vis à vis online platforms. Since there are currently no rules protecting individuals obtaining digital content against counter-performances of non-monetary nature (such as data transfers), this regulatory gap could be interpreted by undertakings as an incentive to move towards business models whose distinctive feature is the supply of digital goods or services without there being any relevant monetary transaction (European Parliament and the Council, 2015, para. 13).

---

[6] This Directive has indeed to be read in conjunction with the guidance provided in 2016 by the Commission on its implementation which recognises that *« data-driven business structures are becoming predominant in the online world. In particular, online platforms analyse, process and sell data related to consumer preferences and other user-generated content. This, together with advertising, often constitutes their main source of revenues»* – COM (2016) 320 final (European Commission, 2016).

In tight relation with such emerging awareness, an example could be provided of how the EU legislator, starting with the 2015 Digital Single Market Strategy, has commenced looking at data in the context of new markets, adjacent to the traditional electronic communications as well as audio-visual media ones, trying to capture brand-new data-related dynamics via tailored soft-touch legislative interventions, with a view not to hampering innovation, while protecting though fundamental goods within the EU legal order.

In this respect, a Directive proposal has been tabled "on certain aspects concerning contracts for the supply of digital content"[7], which covers not only monetary payments, but also payments made in terms of personal or other types of data provided by consumers in exchange for services, establishing that the termination of the contract for lack of conformity of the digital content implies that the supplier shall reimburse the price paid by the consumer or, if this latter's counter-performance consisted in the provision of data, the same supplier shall refrain from using such data and any other information which the consumer has provided in exchange for the digital content (article 13).[8]

This proposal may be a first step in the right direction. However, on the background of the results and considerations contained in this paper, it is reasonable to affirm that policies should first identify what is the "black box" to examine, meaning the expected perimeter of any prospective regulatory intervention in the context of data-related transactions, and then the subjects entrusted to monitor data transactions between individuals and platforms. The idea appears indeed inaccurate that transactional distortions could simply be sorted out through enhanced transparency obligations: such distortions would indeed still feature transactions, as long as strong and structural asymmetric information issues cannot be wiped out by simple transparency rules and individuals are characterized by bounded rationality. The individuals' impulsiveness in the provision of their own data would indeed stay, due to the marked information asymmetry as to such data value and their potential primary and secondary uses, leading individuals not to weigh in costs and benefits associated to data transactions and disregard their consequences. Overall, these mechanisms lead to socially inefficient outcomes, where a disproportionate amount of individual data is used for commercial businesses.

---

[7] With a view to enhancing cross-border trade in the Union, the mentioned proposal couples with the 2017 amended proposal for a Directive on certain aspects concerning contracts for the sales of goods, amending Regulation (EC) No 2006/2004 of the European Parliament and of the Council and Directive 2009/22/EC of the European Parliament and of the Council and repealing Directive 1999/44/EC of the European Parliament and of the Council. Both proposals, on the background of the political agreement reached last January on this whole legislative package, should soon be formally adopted.

[8] Article 16 «also provides that the supplier shall refrain from using data and any other information which the consumer has provided in exchange for the digital content» (European Parliament and the Council, 2015, sec. 13; 16).

# Bibliography

Acquisti, A. (2010). The Economics of Personal Data and the Economics of Privacy. 30 Years after the OECD Privacy Guidelines. *Brussels: OECD*. Retrieved from http://www.oecd.org/internet/ieconomy/theeconomicsofpersonaldataandprivacy30yearsaftertheoecdprivacyguidelines.htm

Acquisti, A., & Grossklags, J. (2005). Privacy and rationality in individual decision making. *IEEE Security and Privacy Magazine*, *3*(1), 26–33. http://doi.org/10.1109/MSP.2005.22

Acquisti, A., & Grossklags, J. (2008a). *Digital privacy : theory, technologies, and practices*. Auerbach Publications.

Acquisti, A., & Grossklags, J. (2008b). What can behavioral economics teach us about privacy. In *Digital privacy: theory, technologies and practices* (pp. 363–377). Auerbach Publications, Taylor & Francis Group.

Akerlof, G. A. (1970). The Market for "'Lemons'" Quality Uncertainty and the Market Mechanism. *The Quarterly Journal of Economics*, *84*(3), 488. http://doi.org/10.2307/1879431

Akram, U., Hui, P., Khan, M. K., Hashim, M., Qiu, Y., & Zhang, Y. (2017). Online Impulse Buying on "Double Eleven" Shopping Festival: An Empirical Investigation of Utilitarian and Hedonic Motivations. In *International Conference on Management Science and Engineering Management* (pp. 680–692). http://doi.org/10.1007/978-3-319-59280-0_56

Anderson, C. (2006). *The long tail: Why the future of business is selling less of more*. Hachette Books.

Arrow, K. J. (1958). Utilities, attitudes, choices: A review note. *Econometrica: Journal of the Econometric Society*, 1–23. http://doi.org/10.2307/1907381

Bayley, G., & Nancarrow, C. (1998). Impulse purchasing: a qualitative exploration of the phenomenon. *Qualitative Market Research: An International Journal*, *1*(2), 99–114. http://doi.org/10.1108/13522759810214271

Benartzi, S., & Thaler, R. H. (1995). Myopic loss aversion and the equity premium puzzle. *The Quarterly Journal of Economics*, *110*(1), 73–92. http://doi.org/10.2307/2118511

Bennett, C. J. (1995). The political economy of privacy: a review of the literature. *Hackensack, NJ: Center for Social and Legal Research*.

Cohen, J. E. (2000). Privacy, Ideology, and Technology: A Response to Jeffrey Rosen. *Georgetown Law Journal*, *89*.

Davies, S. (1997). Re-engineering the right to privacy: how privacy has been transformed from a right to a commodity. In *In Technology and Privacy: The New Landscape, eds P. Agre and M. Rotenberg*. Cambridge, MA: MIT Press.

Delmastro, M., & Nicita, A. (2019). *Big Data*. Il Mulino.

European Commission. Commission Staff Working Document [SWD(2016) 163 final] "Guidance on the Implementation/Application of Directive 2005/29/EC on Unfair Commercial Practices" accompanying Communication from the Commission COM(2016) 320 final (2016). Retrieved from https://eur-lex.europa.eu/legal-content/EN/TXT/?uri=CELEX%3A52016SC0163

European Council. Council Directive 93/13/EEC of 5 April 1993 on unfair terms in consumer contracts (1993). Retrieved from https://eur-lex.europa.eu/legal-content/EN/TXT/PDF/?uri=CELEX:31993L0013&from=EN

European Parliament and European Council. Directive 2011/83/EU of the European Parliament and of the Council of 25 October 2011 on consumer rights, amending Council Directive 93/13/EEC and Directive 1999/44/EC of the European Parliament and of the Council and repealing Council Directive 85/577/EEC and Directive 97/7/EC of the European Parliament and of the Council (2011). Retrieved from https://eur-lex.europa.eu/legal-content/EN/TXT/?uri=celex%3A32011L0083

European Parliament and of the Council. Directive 2005/29/EC of the European Parliament and of the Council of 11 May 2005 concerning unfair business-to-consumer commercial practices in the internal market (2005). Retrieved from https://eur-lex.europa.eu/legal-content/EN/TXT/?uri=celex%3A32005L0029

European Parliament and the Council. Proposal for a Directive of the European Parliament and of the Council [COM(2015) 634 final] "on certain aspects concerning contracts for the supply of digital content" (2015). Retrieved from https://eur-lex.europa.eu/legal-content/EN/TXT/?uri=celex%3A52015PC0634

Kahneman, D. (2011). Thinking fast and slow. Allen Lane. Penguin.

Kahneman, D., & Tversky, A. (1979). Prospect Theory: An Analysis of Decision under Risk. *Econometrica*, *47*(2), 263–292. http://doi.org/10.2307/1914185

Kahneman, D., & Tversky, A. (1986). Rational choice and the framing of decisions. *Journal of Business*, *59*(4), 251–278. http://doi.org/10.2307/2352759

Moe, W. W. (2003). Buying, searching, or browsing: Differentiating between online shoppers using in-store navigational clickstream. *Journal of Consumer Psychology*, *13*(1–2), 29–39. http://doi.org/10.1207/S15327663JCP13-1&2_03

Rook, D. W., & Fisher, R. J. (1995). Normative influences on impulsive buying behavior. *Journal of Consumer Research*, *22*(3), 305–313. http://doi.org/10.1086/209452

Rook, D. W., & Hoch, S. J. (1985). Consuming impulses. *ACR North American Advances*.

Simon, H. (1955). A behavioral model of rational choice. *Quarterly Journal of Economics*, *69*(1), 99–118. http://doi.org/10.2307/1884852

Stern, H. (1962). The significance of impulse buying today. *Journal of Marketing*, *26*(2), 59–62. http://doi.org/10.1177/002224296202600212

Tirole, J. (2017). *Economics for the common good*. Princeton University Press.

Tversky, A., & Kahneman, D. (1991). Loss aversion in riskless choice: A reference-dependent model. *The Quarterly Journal of Economics*, *106*(4), 1039–1061. http://doi.org/10.2307/2937956